\documentclass{llncs}
\usepackage{paralist} \setdefaultitem{}{\textbullet }{ $\star$}{}
\usepackage{url}
\usepackage{setspace}
\usepackage{graphicx}

\begin{document}

\title{Education in Conflict Zones: a Web and Mobility Approach}




\author{Shah Mahmood \inst{1} \and Ismatullah Nazar \inst{2}} 
\institute{Department of Computer Science, University College London,\\
       United Kingdom \\
\and Brunel Business School,  Brunel University, \\ United Kingdom \\
     \email{shah.mahmood@cs.ucl.ac.uk, cb10inn@brunel.ac.uk}}


\maketitle
\begin{abstract}

We propose a new framework for education in conflict zones,
considering the explosive growth of social media, web
services, and mobile Internet over the past decade. Moreover, we focus on one
conflict zone, Afghanistan, as a case study, because of its alarmingly high
illiteracy rate, lack of qualified teachers, rough terrain,
and relatively high mobile penetration of over 50\%. In several of
Afghanistan's provinces, it
is hard to currently sustain the traditional bricks-and-mortar school model, due to
numerous incidents of schools, teachers, and students being attacked because of the ongoing insurgency and political
instability. Our model improves the virtual school
model, by addressing most of its disadvantages, to provide students in
Afghanistan with an opportunity to achieve standardised education, even
when the security situation does not allow them to
attend traditional schools. One of the biggest advantages of this
model is that it is sufficiently robust to deal with gender
discrimination, imposed by culture or politics of the region.

\end{abstract}
\keywords{Education, Conflict Zone, Social Network, Mobile Internet, Afghanistan}

\section{Introduction} \label{sec:Introduction}

\begin{quote}\centering{``Education is a better safeguard of liberty than a
  standing army.'' $\sim$ Edward Everett} \end{quote} 

The explosive rise in the use of web services has redefined human
connectivity in a very significant way. Its influence is evident from the
fact that Facebook has registered over 800 million active users \cite{FacebookStatistics12},
Qzone, the Chinese social network, over 450 million  \cite{Qzone12}, and Twitter over 300
million \cite{Twitter11}. Every month, over 241 million users play social games developed by
Zynga  \cite{Zynga12}. Every day, over 3 billion videos are watched on Youtube and
another 48 hours of videos are uploaded every minute
\cite{YoutubeStatistics12}. Almost 1.2 billion users are currently using
web services over their smart phones, tablets, netbooks, and laptops. On
average, 45\% of the Internet users are below the
age of 25 \cite{MobileStatistics11}. 

This unprecedented connectivity and enormous amount of flow of
information presents untold opportunity to improve human quality of
life worldwide. One such positive
use that we focus on, is to aid learning and provide
more people with opportunities for education. Presently students use
web services including
Google searches, forum postings, and social network discussion boards
to find answers to their queries. The increasing number of video
tutorials have prompted the use of the term Youtube University
\cite{Cohen07}. Although these examples show that the web services are
being used to get help with education, but there is no current
framework which utilises web services to provide education in conflict
zones. Conflicts zones may include places where seeking education by going
to a physical school could
risk the lives of students. They could step on explosive devices, get
shot in the firefight between insurgents and the government, or be
explicitly targeted solely for seeking
a type of education which does not conform with the views of
particular extremist groups.  One such unfortunate country where
students and teachers have been reportedly harmed and schools blown
up, either by the Taliban \cite{Parmer09}, other insurgent groups or 
contractors (Contractors blow up poorly constructed
unfurnished schools
with no equipment to avoid audit and receive full
amount of their contract money).

In this paper, we focus on providing a framework for education in
conflict zones, by considering Afghanistan as a model and
utilising web and telecommunication services . 

The rest of the paper is arranged as follows. First, we provide a background on
the negative impact, on education, of Afghanistan's three decades of
war.  We also discuss the success of mobile networks in the
country, where 50\% of the country's population registered for a
mobile connection within the first 8 years (further details are given in Section
\ref{sec:Background}). Second, we provide details of the proposed
framework for education in conflict zones. This framework is not meant
to completely replace traditional schooling. It is meant to cover for
regions where traditional schooling is not feasible due to security or
other reasons. Our framework includes the use of mobile Internet, social
networks, voice recognition and games (further details about the framework are given in
Section \ref{sec:Framework}). Third, we provide the framework's advantages, including the
  creation of better education opportunities for both female and male
  students, quality assurance, resistance to sabotage,
  transparency and much more (further details in Section
  \ref{sec:Advantages}). Fourth, we discuss some limitations of the
  work and provide possible corresponding solutions
  (further details in Section \ref{sec:Limitations}).  We provide an
  account of related work in Section \ref{sec:RelatedWork} and conclude
  the paper in Section \ref{sec:Conclusion}. 

\section{Background} \label{sec:Background}

In this section, we provide a short overview of the wars in Afghanistan and their
impact on education. Following this we discuss the success of mobile
networks in presence of the ongoing insurgency. 

\paragraph{Afghan wars and their impact on education.}

The Soviet Union invaded Afghanistan on December 27, 1979 by
killing the country's President, replacing him with a Soviet
client, and marching the Soviet army across the border into Afghanistan. This started a bloody insurgency which continues to this
day. Till present, over 2 million Afghans have lost their lives, over 1.5 million
sustained serious injuries, over 1 million became widows, and over 3 millions
orphans \cite{Warloss12,Hilali05}.  These wars also
destroyed the infrastructure of the country including its healthcare system,
security institutions, and education system. According to the United
Nations Development Programme, over 68\% of the
country's population is under the age of 25 \cite{UNDP08}. The
alarming rate of illiteracy in the Southern and Eastern provinces is
evident from the mere fact that Uruzgan's, overall literacy rate is 5\%,
with 10\% men and 0\% women being literate \cite{AusAID12}.  After 2001, there has been some
attempts to educate Afghanistan by forming a renewed curriculum,
building schools, hiring new teachers and providing
students with books and basic stationery. In several parts of the
country, especially South and the East, numerous incidents of schools
being blown up or burnt are reported. Such destruction is either done
by extremist insurgents or contractors for school building. Moreover, teachers have
been kidnapped, bodily harmed and even killed for their support for
education \cite{Parmer09}. Children in such parts are mostly strangers to
education despite some efforts by the global community, such as, 50
schools being built by Canada, in Kandahar province\cite{Canada10}. Even in the
relatively peaceful parts of Afghanistan, teaching standards are
low, due to the unavailability of qualified teachers. Students have to
study by sitting on the floor because the schools are missing
furniture. Teachers are not always present for their classes, busy
working in their fields or running their small businesses
\cite{Burde10}. There is little proper oversight. The
situation is challenging for anyone attempting
to provide quality education using traditional schooling.  

\paragraph{Emerging mobile connectivity.}
Despite all problems in Afghanistan, the mobile communication industry has seen
an impressive growth, since it first started functioning in
2002. According to Afghanistan Telecom Regulatory Authority, the penetration rate of Afghan mobile
market was approximately 50\%, with total users exceeding 12.5 million
\cite{ATRA10}, as of May 2010. The use of mobile phones in Afghanistan
is not limited to communication. Afghanistan's rough and complex
terrain has prompted the use of mobile phones for money transfer. Mobile payments were first launched by Afghan telecom group
Roshan, in partnership with Vodafone.  Though it was initially used
to pay the salaries of police officers in distant areas, its use has
expanded to almost any type of peer to peer money transfers
\cite{MPaisa12}. 

\begin{figure}
 \centering
 \includegraphics[scale=0.75]{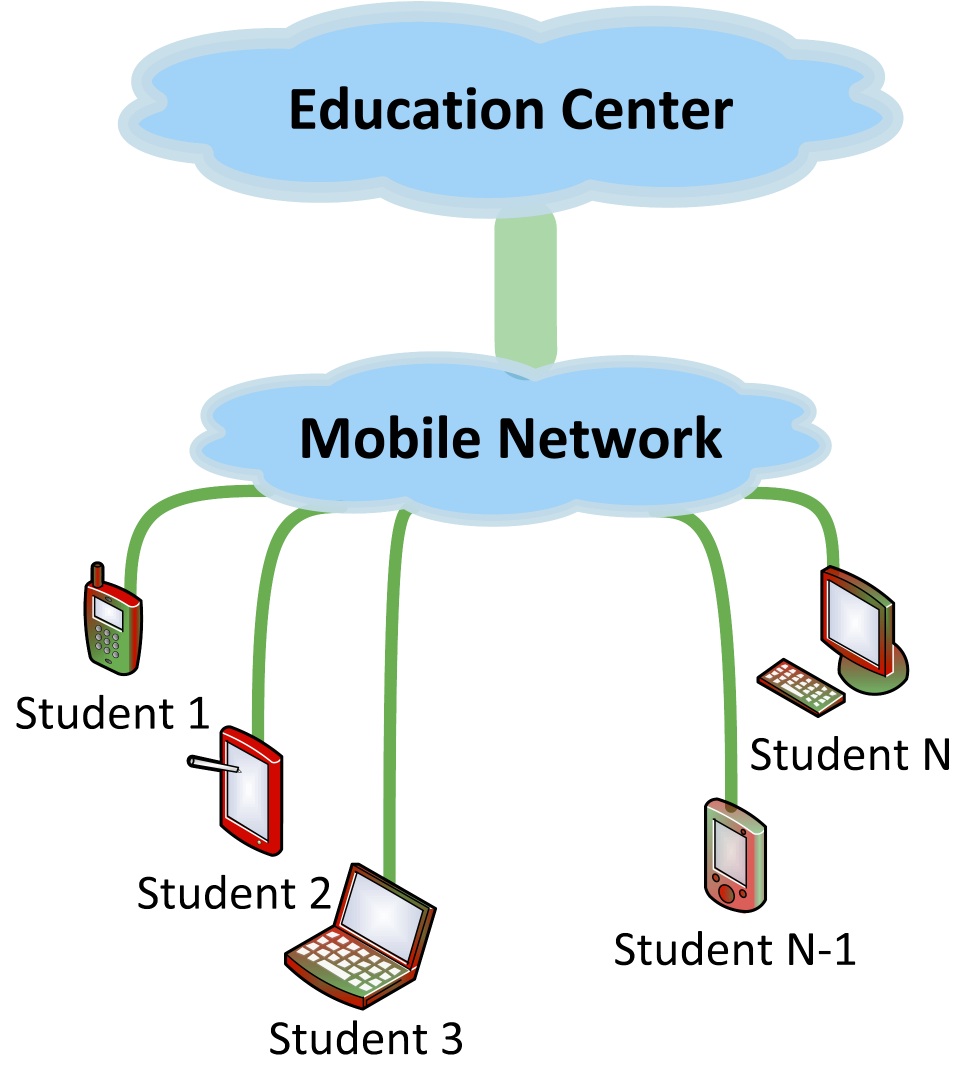}
\caption{Proposed Education Framework}\label{fig:Framework}
\end{figure}

\section{Proposed Framework} \label{sec:Framework}
It is evidently difficult for interested organisations to maintain traditional
schools in conflict zones, when insurgents are consistently destroying
schools at a sustained rate. To counter the problem, we
propose the use of a new education framework, which includes 
elements of both traditional and virtual schooling systems.

The proposed framework, as shown in Figure \ref{fig:Framework},
comprises of three main parts. The first component is the ``Education
Center'' which registers and enrolls all students, physically, by
mail, or online. It also prepares
interactive lectures which are then uploaded into the ``Mobile
Network'' so that they could be downloaded by ``Students''. Each
student is required to have a tablet or any other computing device
with mobile connectivity. These devices could be provided by the
government or education agencies to students from families
having lower incomes.  Lectures should be uploaded periodically, i.e., every few days of
every week, by the education center. Students should be
required to take tests and complete assignments by their
deadlines. Quizzes can be monitored by sending the video stream from the
inbuilt or attached webcam of the 
student's computing device, to the education center. This will prevent any unfair means being
adopted and will also give the education center more data about a
particular student's behaviour during a quiz. Such behavioural data
accumulated over time can enable the education center to provide each
student with targeted material, tailored individually, to enhance
each students learning experience. Other
features of the framework are discussed below.  

\paragraph{Teacher-student interactive video sessions.} It will not be
fair to expect all students to understand everything in the video
lectures. They may have some questions which should be answered after
being posted on the web page for the relevant
course. Each course could be assigned several secondary teachers or
teaching assistants, who should have 1-2 hours of interaction with a
decent sized group of students. During these sessions,
teachers should answer all posted questions, and if time permits, ask
students for other queries and give them tips for learning. These
interactive sessions could be established through video conferencing
between the education center and each students computing
device. Teachers should also be able to see all their students
in order to create an interpersonal bonding. Primary school students who are not able to write
questions may express their lack of understanding for the topic by
pressing the corresponding button at the end of each sub-topic, which
should then be relayed to the teacher and addressed during these
sessions. In some cases, teachers may have individual sessions through
video conferencing with students. 

\paragraph{Use of voice recognition.} Voice recognition can be used for language classes and early
learners. Suppose a child is learning the English alphabet and is
hearing a character `A', the program should pause for the
child's feedback and correct him by repeating the correct
pronunciation. For natives of certain other languages, pronunciation might
not work exactly like natives of the language that they want to learn, so
the software program should adapt to such requirements and constraints. 
This use of technology will save teachers' time and provide each student
with the flexibility of attempting to polish their reading and
speaking skills with the help of software tools. 

\paragraph{Use of games.} About 45\% of British children under 12 have used
Facebook \cite{Barnett12}. They use the social networking site only
to play online games. The popularity of games could be used for education. Quizzes, reading and
writing assignments can be made much more interesting by turning them
into games. Violent games have been reported to negatively affect the
personality of children \cite{Anderson01}, but educational games with social connections could reverse the
negativities and do much more good. Students from conflict zone could
use these games to do group assignments and develop their group
leadership and management skills.

\paragraph{Use of social networks.} Students and their parents should
be automatically registered on the educational social networks. Each
student should be added to the groups corresponding to his courses.
Parents should be added to parents groups. This way students would
stay in touch with their peers and create the social bond which is
otherwise created in traditional schools. Most students' parents in
countries like Afghanistan are not literate.  They can be given the
option to record their messages in voice for postings on the parents'
social group. Parents, using these voice messages, can discuss their child's academic problems
and benefit from the feedback they receive from others in the group.
This will improve the social bonds between parents enabling them to
collectively decide about potentially beneficial decisions for their
children's future. 

Students could be further encouraged to share their academic work
through social networks. The shared material could be a learning
experience for their classmates or others interested, especially those
taking the course next season. The opportunity to comment will
liberate young minds; this way, the Internet can provide the freedom
of speech often sorely lacking in conflict ones. To avoid bullying or
other potential harm, students should be allowed to comment under
pseudonyms or anonymously. As, the groups are supposed to be administered
by the education center, potential for misuse of these pseudonyms is
minimal. 

Young student should be provided adequate protection over the
Internet, by limiting their access to any potentially harmful content.
Unauthorised members should not be permitted to join student groups.

\paragraph{Strict deadlines and time management.} 
As the targeted group to be educated ranges from primary school
students to university level, they can not be expected to manage
their time and complete their tasks without any reminders. Parents of
students should be reminded over their phones through voice and text messages about the upcoming deadlines for their child's school
work. Lack of discipline in time management repeated several times, due
to the negligence of parents, 
should be followed by confidentially requesting a respectful figure in the
neighbourhood to intervene by
advising the negligent parents about the potential harm
they may be doing to their children's future. If neighbours are not
successful in convincing the negligent parents, the problem must be
forwarded to the local education enforcing government authority or a
respectful village chieftain. Peer pressure could be a
good motivator to remind irresponsible parents about their duties.

\paragraph{Creating incentives for hard work.} \label{subsec:Incentives}
Parents of students should be regularly sent mobile messages about
their children's progress. Incentives and constructive competition
should be created between students by announcing best performers
within a locality, only when such announcements are considered safe. The announcement should reach all students and
their parents. Best students for a year should be awarded by sending
them trophies or giving them other types of recognition. Teachers should call high achieving students and their parents to
congratulate them and to ensure that they carry on with their
hard work. 

\paragraph{Induction workshops for students and their parents.}  
Students throughout a conflict zone should be divided into small
manageable localities. Students and parents should be provided
with a induction workshop in their locality by educational organisations
either through physical presence or through a video
conference. Students and their parents should be fully aware
of how to use the tablets or other computing devices and how to
download and play the course material. User interface for such
programs should be suitable for the respective users, i.e., in regions
with higher illiteracy rates, the interface should be more graphical
and symbol oriented, instead of textual. 

\paragraph{Technical support.} 
With technical equipment, we will have technical problems. Students
may break their tablets or the connection might not be working for
countless reasons. Such problems should be addressed by a
technical support hotline within each province or countrywide. This
could be further improved with physical support being available within every few
localities. As most people in Afghanistan live in combined family with
several children in every house, the breakdown of one piece of
equipment will not do much harm. A student with a broken device
should use their sibling's or cousin's device till their own gets
fixed.

\section{Advantages} \label{sec:Advantages}
Advantages of the proposed framework over the current adoption of
traditional school model, in conflict zones like\\ Afghanistan, and
other types of
virtual school models, in other places, are given as follows. 

\paragraph{Equal opportunities for female and male education.} In
places like Afghanistan, over the past few decades, it has been hard to promote education
for girls and women due to intimidation from religious fanatics. These problems are coupled due to cultural constraints. It is not considered a virtuous practice for
females to leave their homes, even for the sole aim of
education. In Afghanistan, there is no cultural/ religious opposition
to girls and women education, if the process does not involve
getting in contact with marriageable men other than one's husband.  As the model uses interactive learning through videos,
social networks, games, etc., it should not create any opposition
either from the hardline religious leaders or the strict followers of
cultural norms. Even if some fanatics have a problem with the
fundamental right of getting educated, they would not be able to
enforce their will on people in their houses.  Lectures heard
through headphones and watched on a tablet could remain a private
affair. When such privacy is expected by any parents or students, then it should
be made sure that the achievements, as explained in Section
\ref{subsec:Incentives}, of such students are not publicised to others.

\paragraph{Quality assurance.}

One big problem for education in conflict zones is the lack of
availability of qualified teaching staff. Those available generally
are unwilling to travel in a poor security environment. This forces
the hiring of sub-standard teaching staff for most
schools. When teaching staff are not suitable for the job, 
students end up being disheartened. Their parents look into
alternatives and force them into jobs or housework.

With the proposed framework, we can assure that the lectures recorded are
delivered by well qualified teachers and the course material is
same for the entire region. The direct interaction sessions are
also expected to be delivered by well qualified teachers or their
assistants.

\paragraph{Economic advantages.} 

The proposed framework is economically much more viable since it can scale to a large number of students. The basic
interactive software material and lecture videos, once created, could be used by any
number of students throughout an entire country, or an entire region
where a particular language is spoken and understood. For the question
and answer sessions, one teacher or teaching assistant can handle many
more students than in a physically interactive group. Government and
education promoting organisations will not need to invest in printing
books and building expensive traditional schools, which may well be
attacked within days, in areas of active insurgency.

\paragraph{Reachability.}

Countries like Afghanistan, with very rough terrain, are very hard
to completely reach, especially in harsh winters when
roads connecting some of the villages are completely blocked for
months. With the proposed framework, teachers do not need to
physically be present in the locality of their students, moreover,
school supplies need not be shipped on regular bases. Everything could
be monitored and controlled, from easily accessible and convenient
centralised locations. 

\paragraph{Hard to sabotage.}

Insurgents in conflict zones can destroy physical schools and
pause the progress of education for the local students for time
spanning over one or more year (time required rebuild a new
school). As there will be no traditional school
buildings, in high risk areas, insurgents will not be able to do much physical
damage. Sabotaging connectivity for our framework can happen
in two ways. Firstly, insurgents will require to destroy the mobile
network. Insurgents also use the same network for their day to day
operations, so they
can not afford harming the network. Moreover, mobile networks have
enough redundancy to withstand attacks by
insurgents in Afghanistan. Secondly, they may use jammers to temporarily disrupt
the signal, but jammers will expose their location. So, in both cases
insurgents will find it much harder to inflict significant harm. 

\paragraph{Avoiding time wastage.}

In some regions of Afghanistan, students need to travel several hours
to reach their schools \cite{Nordland11}. Students need to travel through rivers on
bridges that are not well maintained or through hard mountainous
terrain. According to Oxfam, some students walk up to three hours to
get to school in Balk province\cite{Oxfam10}. 

\paragraph{Security.}

As discussed in the Introduction of this paper, there has been several
incidents of teachers and student being harmed only because of their
thirst for education \cite{UNHCR10}. Students and teachers were exposed to attacks
due to the known location of schools and well defined periods of
study. With education being available at home, insurgent intimidation
will lessen.

\paragraph{Transparency.}

It is hard to monitor corruption in conflict zones, and
Afghanistan has been no different. Powerful warlords can force school
teachers to award their children the best grades while failing the
children of anyone who every opposed them. Officials at the Ministry
of Higher Education in Afghanistan have been found biased, favouring
students' of their own ethnicity and opposing others, in the allocation
of foreign scholarships \cite{Misdaq10}. With
the proposed framework, there will be a digital record of a students
work throughout the year or longer and any drastic changes for
university entry exams or scholarship allocation could be easily
detected, thus creating more transparency.

\section{Limitations and Solutions } \label{sec:Limitations}
There are several limitations, which need to be addressed in order to
fairly construct the proposed framework. First, the user interface for
the educational kit, including downloading and playing of lectures,
use of educational social networks, establishing the interactive video
conference for question and answer session, etc., needs to be very
simple for early students and their parents, who
may not be well educated. Secondly, the mobile network in Afghanistan have
very limited Internet download rates. This should not be a big problem
for our framework because most of the content will not be on the
Internet but on the local network of the education center. Lectures for each week can be cached
at the mobile network's local base station level for easy and quicker download. Social
networks on the education center network will be designed for use between parents
and students within the conflict zone, so that should also be feasible
with current setup of mobile companies in Afghanistan. Mobile
companies are slowly and gradually improving their Internet bandwidth, so
wider sharing over the Internet should become feasible 
relatively quickly. 

Thirdly, electricity is insufficient and unreliable in most parts of
Afghanistan. Most such areas
have access to generators, powered by oil or solar energy, run for a
limited time each day. These generators or the rarely available electricity could be used
to power up the batteries of student tablets in the same way as they
are used  to recharge the mobile phone batteries for their parents,
presently.    

Fourthly, the Internet contains lots of content that could be harmful for
students, or wasteful, at best. Education center's social
network should be carefully monitored and access to inappropriate
material should be limited by the service providers. 

Fifthly, secondary teachers who help students during teacher-student
interaction sessions must also be well qualified. We, through the
Afghan Students Union UK, were able to convince 50 expatriate Afghan
teachers based
in the Netherlands and over a dozen teachers based in the United
Kingdom, to teach Afghan students through video conferencing. In our
discussion with UCL's Department of Political Science, they mentioned
the possibility of establishing web-based and podcast sessions between
UCL and universities in Afghanistan. They also suggested exchange
programmes of
students and staff and possible joint research.

\section{Related work} \label{sec:RelatedWork}
Massachusetts Institute of Technology, in October 2002, took the
initiative of making all its undergraduate and graduate course
material available on the Internet, under the 
open courseware Project. The material available includes
lecture notes, assignments, course syllabi and past exams
\cite{MIT12}.  Though the step taken by MIT was a revolutionary one,
problems get raised due to the non-interactive nature. There is no
well defined way for students to ask teachers, questions about
the available open courseware material. It is more of an aid to self study or
as an aid to classroom material. In our opinion, it is not a full
package that could be adopted at earlier phases of the educational
cycle, e.g. for primary or secondary schooling. 

Khan Academy, started in 2006, offers free educational videos
accessible to anyone on the Internet on a wide range of subjects \cite{KhanAcademy12}. Like
MIT's Open Course Ware Project, Khan's Academy is an impressive
project with a step towards promoting education, using the emerging
web services. Unfortunately, it has the same problems of being used as
a secondary learning tool, in parallel to proper brick and mortar schooling. Its
non-interactive nature with no strict deadlines, make it hard for
primary school students to completely cover required courses.  

Stanford University offered Computer Science
courses online which were completed by almost 32,000 students. For some of the modules, the included homework,
midterm exams, final exams, and the award of a successful completion
certificate at the end of the course \cite{Guzdial11}. Stanfords
courses were more interactive than those offered by MIT and Khan's
Academy but still targeted students aiming for specialisation in a
particular branch of computer science. Our framework is tailored for
students in conflict zones,with difficult access to quality
traditional schools due to security threats and the lack of qualified
teachers. 
\section{Conclusion} \label{sec:Conclusion}

We propose a new framework for education in conflict zones, where
traditional schooling is not feasible. It will combine the positive
features of both traditional and virtual schools with an interactive
program of video lectures, teacher-student interactive video sessions,
use of voice recognition, social networks, and games for enhanced
learning. Students will be provided with quizzes monitored through
their webcam and with strict deadlines for homework. Parents will
regularly be updated about the status of their children's education
through voice and text messages. This framework will have many
advantages including equal opportunity of education for both male and
female students, measures to assure and maintain education quality,
economic feasibility, reachability, resistance to sabotage, and
transparency. We also discussed some of its limitations and provided
mitigating solutions. This framework is not intended to replace traditional schooling
but to provide an alternative more feasible and sustainable education
framework for conflict zones. 

\section*{Acknowledgements}

The authors would like to thank Ashraf Ghani, former finance minister
of Afghanistan, Clare Lockhart, former UN advisor for Afghanistan,
and Paul
Smith, director of the  British Council in Afghanistan, for
discussions about frameworks for education in Afghanistan.  The
authors would also like to thank Alex Braithwaite, for suggestions about joint
research between UCL Department of Political Science and
Afghan Universities. 
\bibliographystyle{abbrv}
\bibliography{References_Education_Conflict}



\end{document}